\documentclass[aps,preprint,showpacs,preprintnumbers,amsmath,amssymb]{revtex4}
\usepackage{amsmath,mathrsfs,amsbsy,color,graphicx,bm,amsthm,amsfonts}
\usepackage{units}
\usepackage{bbm}
\usepackage{times}
\usepackage{dcolumn}
\usepackage{mathrsfs}
\usepackage{amsmath,amssymb,epsfig}
\usepackage{amsmath}
\newcommand{\udots}{\mathinner{\mskip1mu\raise1pt\vbox{\kern7pt\hbox{.}}
\mskip2mu\raise4pt\hbox{.}\mskip2mu\raise7pt\hbox{.}\mskip1mu}}
\begin{document}
\title{Nonlocal correlation in quantum network under relativistic motion}
\author{Si-Han Li, Tian-Yang Wang, Ai-Yan Tong, Shu-Min Wu\footnote{Email: smwu@lnnu.edu.cn}}
\affiliation{ Department of Physics, Liaoning Normal University, Dalian 116029, China}


\begin{abstract}
We investigate the relativistic dynamics of network nonlocality in general $n$-local networks with chain and star topologies using the Unruh-DeWitt detector model.  We show that the relativistic degradation of network nonlocality is strongly governed by the underlying topology.  While chain networks suffer an irreversible sudden death of non-$n$-locality under relativistic motion, star networks exhibit remarkable resilience against relativistic decoherence. Most strikingly, a minimal star network with three peripheral nodes exhibits a remarkable sudden death-sudden birth transition of network nonlocality as the acceleration increases. This reentrant behavior reveals a dual role of the Unruh effect: it can both suppress and protect network nonlocality, offering a new perspective on the relativistic effects of acceleration on quantum networks.
For larger star networks ($n>3$), non-$n$-local correlations persist over the entire acceleration regime. These insights provide valuable conceptual guidance for the structural optimization and design of acceleration-resilient architectures for future relativistic quantum communication and sensing protocols.
\end{abstract}

\vspace*{0.5cm}
\maketitle
\section{Introduction}
Quantum networks constitute the cornerstone for developing future large-scale quantum communication, distributed quantum computing, and long-baseline quantum sensing \cite{A1,A2,A3,A4,A5}. Unlike traditional bipartite quantum structures, a quantum network comprises multiple independent entanglement sources that distribute quantum states to geographically separated nodes, thereby establishing rich and intricate multipartite correlations \cite{B1,B2}. This unique topological architecture gives rise to network nonlocality, a novel and stronger form of multipartite quantum correlation that transcends the linear characterization of standard Bell inequalities \cite{B1,C1,C2}. To rigorously verify this phenomenon, the framework of $n$-locality inequalities has been established, encompassing various typical network geometries such as chain and star configurations \cite{D1,D2,G1}. The violation of these nonlinear inequalities not only serves as definitive evidence for network nonlocality but also provides profound physical insights into the nonlocal properties dictated by complex topologies \cite{B1,C1,E1,G1}. Within this $n$-locality framework, quantum networks permit the verification of joint multipartite correlations by exploiting the source-independence assumption, thereby demonstrating unprecedented advantages in quantum cryptography and nonlocal optical interferometry \cite{F1,F2,F3}. Consequently, it is of profound theoretical significance to investigate the robustness of such nonlocal network correlations against realistic physical environments, as it offers critical guidance for the real-world deployment of device-independent quantum protocols \cite{F1,F2,G2,G3}.

Relativistic quantum information (RQI), lying at the intersection of quantum information theory, quantum field theory, and general relativity, aims to understand how relativistic motion and spacetime geometry influence fundamental quantum resources, such as entanglement, quantum coherence, and nonlocality \cite{qdb1,qdb2,qdb3,qdb4,qdb5,qdb6,qdb7,qdb8,qdb9,qdb10,qdb11,qdb12,qdb13,qdb14,qdb15,qdb16,qdb17,qdb18,qdb19,qdb20,qdb21,qdb22,qdb23,qdb24,qdb25,qdb26,qdb27,qdb28,qdb29,qdb30,qdb31,qdb32,qdb33}. Beyond deepening our understanding of the interplay between quantum theory and gravity, it also seeks to develop methods for protecting and manipulating quantum resources in relativistic environments and to exploit quantum technologies for probing spacetime structures \cite{qdb34,qdb35,qdb36,qdb37,qdb38,qdb39,qdb40}. Traditionally, studies in RQI have mainly relied on the free-field mode formalism, where relativistic effects are characterized through correlations between globally extended field modes perceived by different observers \cite{qdb1,qdb2,qdb3,qdb4,qdb5,qdb6,qdb7,qdb8,qdb9,qdb10,qdb11,qdb12}. However, this approach is intrinsically limited in describing the interaction between localized quantum systems and quantum fields encountered in realistic quantum measurements. To overcome this limitation, the Unruh-DeWitt detector model has been widely adopted \cite{qdb41,qdb42,qdb43,qdb44,qdb45,qdb46,qdb47,qdb48,qdb49,qdb50}. In this framework, localized two-level quantum systems interact with quantized fields over finite proper-time intervals, providing an operationally well-defined framework for investigating the dynamical evolution of quantum resources under relativistic motion.

Within this framework, a natural yet largely unexplored question is how relativistic motion influences quantum correlations with complex network structures, particularly the interplay between network topology and quantum correlations. Unlike conventional relativistic multipartite quantum states, whose correlations are encoded in a single global wave function, relativistic quantum networks consist of multiple independent entanglement sources connecting spatially separated quantum nodes. Consequently, their correlation structure is determined not only by the underlying entanglement resources but also by network topology and source independence. This modular architecture endows quantum networks with superior scalability and experimental feasibility, while the framework of $n$-locality enables the characterization of network nonlocal correlations beyond conventional Bell nonlocality, making relativistic quantum networks a promising platform for exploring the evolution of complex quantum correlations. Existing studies have shown that relativistic effects generally act as effective thermal decoherence, leading to the degradation of quantum resources and even the irreversible sudden death of entanglement and other quantum correlations \cite{qdb50,SDF61,SDF62,SDF63}. However, for quantum networks possessing nontrivial topologies and multiple independent entanglement sources, it remains an open and fundamentally important question whether network topology can fundamentally alter this conventional decoherence scenario, allowing vanished network nonlocal correlations to recover or even exhibit nontrivial revival phenomena. Addressing this question will not only deepen our understanding of the dynamics of quantum correlations in relativistic settings but also provide theoretical guidance for designing robust relativistic quantum communication and quantum sensing networks.

Based on these motivations, we systematically investigate the relativistic evolution of network nonlocality in both chain and star quantum networks within the framework of the Unruh-DeWitt detector model. We first examine the elementary three-node network and demonstrate that an asymmetric allocation of initial entanglement enhances the robustness of network nonlocality against Unruh-induced decoherence compared with a symmetric resource distribution. We then extend our analysis to general $n$-local networks and show that the network topology plays a decisive role in determining the relativistic behavior of multipartite nonlocal correlations. In particular, chain networks become increasingly fragile as the number of intermediate nodes grows, leading to the irreversible disappearance of non-$n$-local correlations, whereas star networks exhibit significantly stronger robustness against relativistic degradation. Remarkably, for the minimal star network ($n=3$), the vanished network nonlocality reappears in the infinite-acceleration limit, while for larger star networks ($n>3$), non-$n$-local correlations survive throughout the entire acceleration regime. These results reveal a topology-dependent mechanism for protecting network nonlocality under relativistic motion and provide new insights into the design of robust relativistic quantum communication and quantum sensing networks.

The remainder of this paper is organized as follows. In Sec.~II, we introduce the Unruh-DeWitt detector model and derive the relativistic evolution of the detector states. In Sec.~III, we briefly review the theoretical framework of quantum network nonlocality and the corresponding $n$-locality inequalities. In Sec.~IV, we investigate the relativistic evolution of non-bilocal correlations in the elementary three-node quantum network. In Sec.~V, we extend the analysis to general $n$-local quantum networks and examine the impact of network topology on relativistic network nonlocality. Finally, Sec.~VI presents our conclusions.

\section{Evolution of the detectors' state under Unruh-DeWitt detector model}
\begin{figure}
\centering
\includegraphics[width=0.86\textwidth]{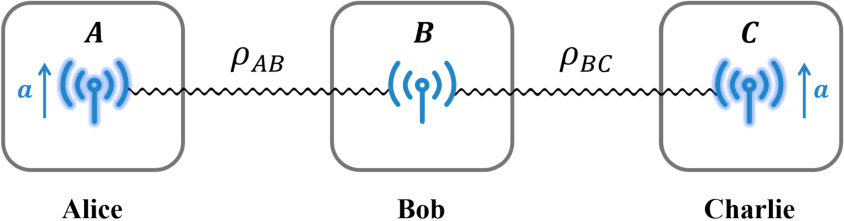}
\caption{Schematic illustration of the elementary bilocal quantum network in the Unruh-DeWitt detector model. Two independent entanglement sources distribute entangled detector pairs to Alice-Bob and
Bob-Charlie, respectively. The Bob node therefore receives one detector from each source. The detectors associated with Alice and Charlie undergo uniform acceleration $a$ during a finite interaction time $\Delta$, whereas Bob's detectors remain inertial.}
\label{Fig1}
\end{figure}

We consider an elementary bilocal quantum network consisting of three spatially separated observers, Alice ($A$), Bob ($B$), and Charlie ($C$), each equipped with Unruh-DeWitt detectors modeled as noninteracting two-level atoms. Since Bob shares independent entangled states with both Alice and Charlie, the Bob node is composed of two identical inertial detectors, denoted by $B_{1}$ and $B_{2}$, which interact independently with the two entanglement sources. The detectors of Alice and Charlie are switched on and undergo uniform proper acceleration $a$ during a finite proper-time interval $\Delta$, whereas the detectors at the Bob node remain inertial throughout the evolution (see Fig.~\ref{Fig1}). After this interval, the detectors of Alice and Charlie continue their uniform acceleration while remaining coupled to the field. Their worldlines are given by
\begin{equation}\label{w66}
t(\tau)=a^{-1}\sinh a\tau, \quad x(\tau)=a^{-1}\cosh a\tau,
\end{equation}
with $y(\tau)=z(\tau)=0$, where $a$ is the proper acceleration of Alice and Charlie, and $\tau$ represents the proper time of the detectors \cite{SDF61,SDF62}. Throughout this paper, we set $c=\hbar=k_{B}=1$.

We assume that the elementary bilocal network is initially established by two independent entanglement sources, which distribute identical bipartite entangled states to the detector pairs $AB$ and $BC$, respectively. The initial states are
\begin{equation}\label{qq7}
|\Psi_{AB}\rangle=\sin{\theta_1}|01\rangle+\cos{\theta_1}|10\rangle,
\end{equation}
\begin{equation}\label{ql7}
|\Psi_{BC}\rangle=\sin{\theta_2}|01\rangle+\cos{\theta_2}|10\rangle.
\end{equation}
Since the dynamical evolution of the subsystems $AB$ and $BC$ is completely analogous, we focus on the subsystem $AB$ in the main text, while the analysis for $BC$ is presented in Appendix~A. Accordingly, the initial state of the detector--field system $AB\phi$ at time $t_{0}$ is

\begin{equation}\label{A7}
|\Psi^{AB\phi}_{t_{0}}\rangle=|\Psi_{AB}\rangle\otimes|0_{M}\rangle,
\end{equation}
where $|0_{M}\rangle$ denotes the Minkowski vacuum of the scalar field, and $|0\rangle$ and $|1\rangle$ represent the ground and excited states of each detector, respectively.

The total Hamiltonian of system $AB\phi$ is given by
\begin{equation}\label{A9}
H_{AB\phi}=H_{A}+H_{B}+H_{KG}+H_{\rm{int}}^{A\phi},
\end{equation}
where $H_{T} = \Omega T^{\dagger}T$ ($T=A,B$) denotes the free Hamiltonian of the detectors with energy gap $\Omega$, $H_{KG}$ is the Hamiltonian of the massless scalar field, and $H_{\rm{int}}^{A\phi}$ describes the coupling between Alice's detector and the field. The ladder operators satisfy $T|1\rangle=|0\rangle$, $T^{\dagger}|0\rangle=|1\rangle$, and $T^{\dagger}|1\rangle=T|0\rangle=0$. We model the interaction by the usual Unruh-DeWitt coupling localized around the detector $A$
\begin{equation}\label{A8}
H_{\rm{int}}^{A\phi}(t)=\epsilon(t)\int_{\Sigma_{t}}d^{3} \mathbf{x}\sqrt{-g}\phi(x)[\chi(\mathbf{x})A+\bar{\chi}(\mathbf{x})A^{\dagger}],
\end{equation}
where $g\equiv \det(g_{ab})$ and $g_{ab}$ is the Minkowski spacetime metric. The function $\chi(\mathbf{x})=(\kappa\sqrt{2\pi})^{-3}\exp(-\mathbf{x}^{2}/(2\kappa^{2}))$ is a Gaussian coupling function, where the parameter $\kappa$ governs the effective size or interaction range of the detector. This profile vanishes outside a small region around the detector \cite{SDF61}, effectively modeling a pointlike detector interacting only with nearby field modes in the Minkowski vacuum.

In the weak-coupling regime, the final state $|\Psi^{AB\phi}_{t=t_{0}+\Delta}\rangle$ of the atom-field system at time $t=t_{0}+\Delta$ can be calculated in first-order perturbation theory over the coupling constant $\epsilon$. Under the evolution generated by the Hamiltonian in Eq.(\ref{A9}), the final state $|\Psi^{AB\phi}_{t}\rangle$ at time $t$ takes the form 
\begin{equation}\label{qq1}
|\Psi^{AB\phi}_{t}\rangle=\{I-i[\phi(f)A+\phi(f)^{\dagger}A^{\dagger}]\}|\Psi^{AB\phi}_{t_{0}}\rangle.
\end{equation}
In this expression, the smeared field operator $\phi(f)$ is defined as 
\begin{equation}\label{qq2}
\begin{split}
\phi(f)\equiv&\int d^{4}x\sqrt{-g}\chi(x)f\\
=&i[a_{RI}(\overline{uE\overline{f}})-a^{\dagger}_{RI}(uEf)],
\end{split}
\end{equation}
and describes the distribution of the external scalar field. Here, $f\equiv\epsilon(t)e^{-i\Omega t}\chi(\mathbf{x})$ is a complex function with compact support in Minkowski spacetime, while $a_{RI}(\overline{u})$ and $a_{RI}^{\dagger}(u)$ represent the annihilation and creation operators of $u$ modes, respectively. The operator $u$ extracts the positive-frequency components of Klein-Gordon solutions in the Rindler metric, and $E$ is the difference between the advanced and retarded Green's functions.

Although the two entangled pairs are initially prepared in the same state, the Unruh effect acts on different detectors in the two subsystems, namely Alice in the subsystem $AB$ and Charlie in the subsystem $BC$. Consequently, the accelerated detector occupies different positions within each bipartite state, giving rise to distinct relativistic evolutions. Since the two cases can be treated analogously, we present only the evolution of the subsystem $AB$ in the main text. Substituting the initial state in Eq.(\ref{A7}) into Eq.(\ref{qq1}), we obtain the evolved state for system $AB$ in terms of the Rindler operators $a^{\dagger}_{RI}$ and $a_{RI}$ as 
\begin{equation}\label{qqq2}
|\Psi^{AB\phi}_{t}\rangle=|\Psi^{AB\phi}_{t_{0}}\rangle+\cos{\theta_1}|00\rangle\otimes \big[a^{\dagger}_{RI}(\lambda)|0_{M}\rangle \big]+\sin{\theta_1}|11\rangle\otimes \big[a_{RI}(\bar{\lambda})|0_{M}\rangle \big].
\end{equation}
Here, $\lambda=-uEf$, and the Rindler operators $a^{\dagger}_{RI}(\lambda)$ and $a_{RI}(\overline{\lambda})$ are defined in Rindler region $I$. The Minkowski vacuum is denoted by $|0_{M}\rangle$. The Bogoliubov transformations between the Rindler operators and the operators annihilating the Minkowski vacuum state are given by
\begin{equation}\label{qq3}
a_{RI}(\bar{\lambda})=\frac{a_{M}(\overline{F_{1\Omega}})+e^{-\pi\Omega/a}a^{\dagger}_{M}(F_{2\Omega})}{(1-e^{-2\pi\Omega/a})^{1/2}},
\end{equation}
\begin{equation}\label{qq5}
a^{\dagger}_{RI}(\lambda)=\frac{a^{\dagger}_{M}(F_{1\Omega})+e^{-\pi\Omega/a}a_{M}(\overline{F_{2\Omega}})}{(1-e^{-2\pi\Omega/a})^{1/2}},
\end{equation}
where $F_{1\Omega}=\frac{\lambda+e^{-\pi\Omega/a}\lambda\circ w}{(1-e^{-2\pi\Omega/a})^{1/2}}$ and $F_{2\Omega}=\frac{\overline{\lambda\circ w}+e^{-\pi\Omega/a}\bar{\lambda}}{(1-e^{-2\pi\Omega/a})^{1/2}}$ \cite{SDF62,SDF63}. Here, the wedge reflection isometry $w(t,x,y,z)=(-t,-x,y,z)$ maps points in the Rindler region $I$ to region $II$, reflecting $\lambda$ to $\lambda\circ w$.

By applying the Bogoliubov transformations in Eqs.(\ref{qq3}) and (\ref{qq5}) along with the relations $a_{M}|0_{M}\rangle=0$ and $a_{M}^{\dagger}|0_{M}\rangle=|1_{M}\rangle$, the evolved state in Eq.(\ref{qqq2}) can be recast in the form 
\begin{equation}\label{qqq5}
|\Psi^{AB\phi}_{t}\rangle=|\Psi^{AB\phi}_{t_{0}}\rangle+\nu \left [ \cos{\theta_1}\frac{|00\rangle
\otimes|1_{\tilde{F}_{1\Omega}}\rangle}{(1-e^{-2\pi\Omega/a})^{1/2}} + e^{-\pi\Omega/a} \sin{\theta_1} \frac{|11\rangle\otimes|1_{\tilde{F}_{2\Omega}}\rangle}{(1-e^{-2\pi\Omega/a})^{1/2}} \right],
\end{equation}
where $\tilde{F}_{i\Omega}=F_{i\Omega}/\nu$ ($i=1,2$). To determine the state of the detectors after their interaction with the field, we trace out the external field degrees of freedom. This yields the reduced density matrix of the detector subsystem $AB$,
\begin{equation}
\rho^{{AB}}_{t}=\|\Psi^{AB\phi}_{t}\|^{-2}\mathrm{Tr}_{\phi}|\Psi^{AB\phi}_{t}\rangle\langle \Psi^{AB\phi}_{t}|,
\end{equation}
where the normalization factor $\|\Psi^{AB\phi}_{t}\|^{2}$ has the form
\[
\|\Psi^{AB\phi}_{t}\|^{2}=1+\frac{\nu^{2}(e^{-2\pi\Omega/a}\sin^2\theta_1+\cos^2\theta_1)}{1-e^{-2\pi\Omega/a}}.
\]
Accordingly, the reduced density matrix of the detectors takes the form
\begin{equation}\label{pp6}
\rho^{{AB}}_{t}=
 \left(\!\!\begin{array}{cccc}
\gamma_1 & 0 & 0 & 0 \\
0 & 2 \alpha_1 \sin^2\theta_1 & \alpha_1 \sin2\theta_1 & 0 \\
0 & \alpha_1 \sin2\theta_1 & 2 \alpha_1 \cos^2\theta_1 & 0 \\
0 & 0 & 0 & \beta_1
\end{array}\!\!\right),
\end{equation}
where the matrix elements $\alpha_1$, $\beta_1$, and $\gamma_1$ are given by 
\begin{equation}\label{l6}
\alpha_1=\frac{1-q}{2(1-q)+2\nu^{2}(q \sin^2\theta_1+\cos^2\theta_1)},
\end{equation}
\begin{equation}\label{ll6}
\beta_1=\frac{\nu^{2} q \sin^2\theta_1}{(1-q)+\nu^{2}(q \sin^2\theta_1+\cos^2\theta_1)},
\end{equation}
\begin{equation}\label{lll6}
\gamma_1=\frac{\nu^{2} \cos^2\theta_1}{(1-q)+\nu^{2}(q \sin^2\theta_1+\cos^2\theta_1)},
\end{equation}
and the acceleration parameter $q$ is defined as $q\equiv e^{-2\pi\Omega/a}$. 
The effective coupling strength is defined as $\nu^{2}\equiv\|\lambda\|^{2}=\frac{\epsilon^{2}\Omega\Delta}{2\pi}e^{-\Omega^{2}\kappa^{2}}$. For the validity of this definition, the condition $\Omega^{-1}\ll\Delta$ must hold. The parameter $\nu$ quantifies the effective interaction strength between the detector and the scalar field. Consequently, the condition $\nu^{2}\ll1$ corresponds to the weak-coupling limit, ensuring that the interaction energy is small relative to the intrinsic energy scales of the system. In this regime, the Dyson series expansion of the time-evolution operator allows for a valid truncation at the leading order, rendering higher-order contributions negligible. Thus, to ensure the validity of our perturbative approach, we strictly impose $\nu^{2}\ll1$ throughout the analysis. It is important to note that $q$ is a monotonic function of the acceleration $a$. Specifically, $q\rightarrow0$ and $q\rightarrow1$ correspond to the limits of zero and infinite acceleration, respectively.

Note that any two-qubit state can be transformed into a Bell-diagonal state via appropriate local unitary transformations without changing its nonlocality. Therefore, we consider the initial states to be Bell-diagonal states, which are explicitly represented as \cite{B1}
\begin{equation}\label{LS1}
\rho_{AB} = \frac{1}{4} \left( I_2 \otimes I_2 + \mathbf{a} \cdot \sigma \otimes I_2 + I_2 \otimes \mathbf{b} \cdot \sigma + \sum_{i=1}^3 t_i \sigma_i \otimes \sigma_i \right),
\end{equation}
where $\sigma = (\sigma_1, \sigma_2, \sigma_3)$ with $\sigma_i\ (i=1,2,3)$ denoting the standard Pauli matrices, and $\mathbf{a}, \mathbf{b} \in \mathbb{R}^3$ represent the local Bloch vectors. It is well known that the bipartite $X$-state can be expressed as
\begin{equation}\label{LS2}
\rho^X = \begin{pmatrix} 
\rho_{11} & 0 & 0 & \rho_{14} \\ 
0 & \rho_{22} & \rho_{23} & 0 \\ 
0 & \rho_{23} & \rho_{33} & 0 \\ 
\rho_{14} & 0 & 0 & \rho_{44} 
\end{pmatrix},
\end{equation}
where $\rho_{ij}\ (i,j=1,2,3,4)$ are real parameters. By employing appropriate local unitary transformations, the $X$-state can be rewritten in the Bloch-Fano representation as 
\begin{equation}\label{LS3}
\rho^X = \frac{1}{4} \begin{pmatrix} 
1+t_3+a_3+b_3 & 0 & 0 & t_1-t_2 \\ 
0 & 1-t_3+a_3-b_3 & t_1+t_2 & 0 \\ 
0 & t_1+t_2 & 1-t_3-a_3+b_3 & 0 \\ 
t_1-t_2 & 0 & 0 & 1+t_3-a_3-b_3 
\end{pmatrix},
\end{equation}
with the corresponding Bloch coefficients explicitly given by
\begin{equation}\label{LS4}
\begin{aligned}
& t_1 = 2(\rho_{23} + \rho_{14}), \quad t_2 = 2(\rho_{23} - \rho_{14}), \quad t_3 = \rho_{11} - \rho_{22} - \rho_{33} + \rho_{44}, \\
& a_3 = \rho_{11} + \rho_{22} - \rho_{33} - \rho_{44}, \quad b_3 = \rho_{11} - \rho_{22} + \rho_{33} - \rho_{44}.
\end{aligned}
\end{equation}

From Eq.(\ref{pp6}), it is obvious that the reduced density matrix of the detectors $\rho^{{AB}}_{t}$ satisfies the form of the bipartite $X$-state. Rewriting $\rho^{AB}_{t}$ in the Bloch-Fano decomposition according to Eq.(\ref{LS1}) yields
\begin{equation}\label{LS5}
\begin{split}
\rho^{AB}_{t} =& \frac{1}{4} \Big[ I_2 \otimes I_2 + (\gamma_1 - 2 \alpha_1 \cos2\theta_1 - \beta_1) \sigma_3 \otimes I_2 + (\gamma_1 + 2 \alpha_1 \cos2\theta_1 - \beta_1) I_2 \otimes \sigma_3 \\
& + 2 \alpha_1 \sin2\theta_1 \sigma_1 \otimes \sigma_1 + 2 \alpha_1 \sin2\theta_1 \sigma_2 \otimes \sigma_2 + (\gamma_1 - 2 \alpha_1 + \beta_1) \sigma_3 \otimes \sigma_3 \Big].
\end{split}
\end{equation}
Here, the relevant coefficients are identified as $t_1^{(1)}=t_2^{(1)}=2 \alpha_1 \sin2\theta_1$, $t_3^{(1)}=\gamma_1 - 2 \alpha_1 + \beta_1$, $a_3^{(1)}=\gamma_1 - 2 \alpha_1 \cos2\theta_1 - \beta_1$ and $b_3^{(1)}=\gamma_1 + 2 \alpha_1 \cos2\theta_1 - \beta_1$.

\section{Quantum network nonlocal correlations}
Consider a general network $\Xi_{m,n}$ consisting of $m$ parties, namely, $\mathcal{A}_{1},\mathcal{A}_{2},\dots,\mathcal{A}_{m}$, and $n$ sources $\rho_{1},\rho_{2},\dots,\rho_{n}$. Each party $\mathcal{A}_{i}$ can perform a local measurement labeled by $x_{i}$ ($x_{i}\in \{0, 1\}$) with its respective outcome denoted by $a_{i}$ ($a_{i}\in \{0,1\}$). The behavior of this network is classified as local if its joint probability distribution can be decomposed in the following form:
\begin{equation}\label{eq4}
P(\mathbf{a}|\mathbf{x})=\int_{\Omega}\prod_{i=1}^{n}d\mu_{i}(\lambda_{i})\prod_{j=1}^{m}p(a_{j}|x_{j},\Lambda_{j}),
\end{equation}
where $\mathbf{a}=(a_{1},\dots,a_{m})$, $\mathbf{x}=(x_{1},\dots,x_{m})$, $\lambda_{i}$ represents the hidden variable distributed by source $\rho_{i}$, and $\mu_{i}(\lambda_{i})$ is the normalized probability distribution satisfying $\int\mu_{i}(\lambda_{i})d\lambda_{i}=1$. Here, $\Lambda_{j}$ denotes the set of classical variables associated with party $\mathcal{A}_{j}$. Otherwise, the correlation behavior is defined as network nonlocal \cite{30}.

A network is a $k$-independent network if there are $k$ parties that do not share any sources with each other. Let $\Gamma=\{i_{1},i_{2},\dots,i_{k}\}$ denote the set of indices corresponding to these independent nodes. The joint correlations derived from such a network with classical variables satisfy the following non-linear inequality \cite{30}:
\begin{equation}\label{eq_B_ineq}
\mathcal{B}=|I(m,k)|^{\frac{1}{k}}+|J(m,k)|^{\frac{1}{k}}\le1,
\end{equation}
where the correlators $I(m,k)$ and $J(m,k)$ are defined as
\begin{equation}\label{eq5}
\begin{split}
I(m,k) &= \frac{1}{2^{k}}\sum_{x_{j},j\in\Gamma}\langle A_{x_{1}}A_{x_{2}}\cdots A_{x_{m}}\rangle, \\
J(m,k) &= \frac{1}{2^{k}}\sum_{x_{j},j\in\Gamma}(-1)^{\sum_{j\in\Gamma}x_{j}}\langle A_{x_{1}}A_{x_{2}}\cdots A_{x_{m}}\rangle.
\end{split}
\end{equation}
Herein, $A_{x_{i}}$ represents the observable of the party $\mathcal{A}_{i}$ ($i=1,2,\dots,m$), and the joint expectation value is given by $\langle A_{x_{1}}A_{x_{2}}\cdots A_{x_{m}}\rangle=\sum_{\mathbf{a}}(-1)^{\sum_{j=1}^{m}a_{j}}P(\mathbf{a}|\mathbf{x})$, with $P(\mathbf{a}|\mathbf{x})$ defined in Eq.(\ref{eq4}). This formulation establishes a set of $n$-local nonlinear correlation inequalities based on the assumption of a $k$-independent network. Any violation of Ineq.(\ref{eq_B_ineq}) by a physical system serves as a definitive nonlocality witness for the network configuration under consideration.

\begin{figure}
\centering
\includegraphics[width=0.8\textwidth]{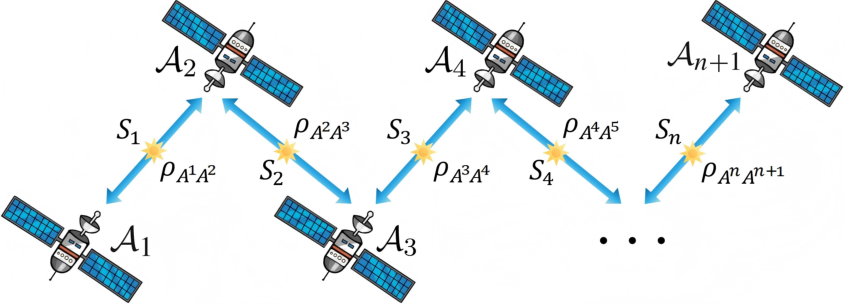}
\caption{The chain network scenario. Here, we choose $\mathcal{A}_1$ and $\mathcal{A}_{n+1}$ as the marginal nodes, and each source $\mathcal{S}_i$ distributes the quantum state $\rho_{A^i A^{i+1}}$ to party $\mathcal{A}_i$ and party $\mathcal{A}_{i+1}$ ($i=1,2 \dots n$).}
\label{Fig2}
\end{figure}

The chain network with $n+1$ parties and $n$ sources shown in Fig.\ref{Fig2} can be modeled as a $2$-independent network. According to Ref. \cite{41}, with respect to the generic quantum state $\rho_{A^{1}A^{2}}\otimes\rho_{A^{2}A^{3}}\otimes\cdots\otimes\rho_{A^{n}A^{n+1}}$, the maximal value of $\mathcal{B}$ in Ineq.(\ref{eq_B_ineq}) is given by
\begin{equation}\label{eq6}
\mathcal{B}_{\rm{chain}}^{\rm{max}}=\sqrt{\prod_{i=1}^{n}\delta_{1}^{(i)}+\prod_{i=1}^{n}\delta_{2}^{(i)}},
\end{equation}
where $\delta_{1}^{(i)}$ and $\delta_{2}^{(i)}$ represent the two largest positive eigenvalues of the matrix $\sqrt{T_{A^{i}A^{i+1}}^{\dagger}T_{A^{i}A^{i+1}}}$ satisfying $\delta_{1}^{(i)}>\delta_{2}^{(i)}$. Here, $T_{A^{i}A^{i+1}}$ denotes the correlation matrix of $\rho_{A^{i}A^{i+1}}$. Notably, in the subsequent discussion of this paper, the correlation matrix of the specific two-qubit state $\rho_{AB}$ is denoted by $T_{AB}=(t_{ij})$, where $i, j\in \{x,y,z\}$ and the matrix elements are defined via $t_{ij}=\mathrm{Tr}(\rho_{AB}\sigma_{i}\otimes\sigma_{j})$.

\begin{figure}
\centering
\includegraphics[width=0.45\textwidth]{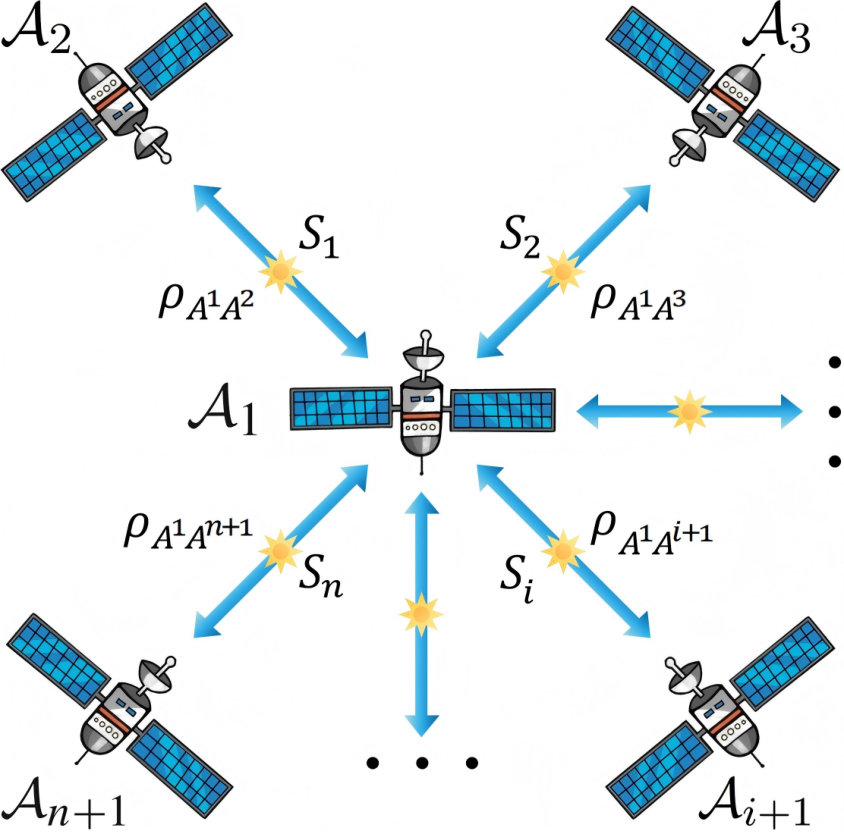}
\caption{The star network scenario. Here, we choose $\mathcal{A}_1$ as the intermediate node, and each source $\mathcal{S}_i$ distributes the quantum state $\rho_{A^1 A^{i+1}}$ to party $\mathcal{A}_1$ and party $\mathcal{A}_{i+1}$ ($i=1,2 \dots n$).}
\label{Fig3}
\end{figure}

Concurrently, the star network comprising $n+1$ parties and $n$ sources shown in Fig.\ref{Fig3} can be considered as an $n$-independent network. According to Ref. \cite{42}, with respect to the generic quantum state $\rho_{A^{1}A^{2}}\otimes\rho_{A^{1}A^{3}}\otimes\cdots\otimes\rho_{A^{1}A^{n+1}}$, the maximal value of $\mathcal{B}$ is given by
\begin{equation}\label{eq7}
\mathcal{B}_{\rm{star}}^{\rm{max}}=\sqrt{\prod_{i=1}^{n}(\delta_{1}^{(i)})^{\frac{2}{n}}+\prod_{i=1}^{n}(\delta_{2}^{(i)})^{\frac{2}{n}}}.
\end{equation}
In this scenario, $\delta_{1}^{(i)}$ and $\delta_{2}^{(i)}$ correspond to the two largest eigenvalues of the matrix $\sqrt{T_{A^{1}A^{i+1}}^{\dagger}T_{A^{1}A^{i+1}}}$ with $\delta_{1}^{(i)}>\delta_{2}^{(i)}$, where $T_{A^{1}A^{i+1}}$ represents the correlation matrix of the state $\rho_{A^{1}A^{i+1}}$.

\section{Evolution of non-bilocality in the ES network}
We start with the elementary entanglement swapping (ES) network consisting of three spatially separated nodes $A$, $B$, and $C$ (corresponding to Alice, Bob, and Charlie in Sec.~II, respectively), characterized by the network state $\rho_{AB}\otimes\rho_{BC}$. This scenario corresponds to the simplest form of both the chain network (Fig.~\ref{Fig2}) and the star network (Fig.~\ref{Fig3}). We refer to the nonlocal correlations generated across this three-node network as nonbilocal correlations or non-bilocality.
Initially, two independent entanglement sources distribute bipartite entangled states, given by Eqs.~\eqref{qq7} and \eqref{ql7}, to the detector pairs $A$-$B$ and $B$-$C$, respectively.
We investigate the specific configuration where the two marginal nodes, $A$ and $C$, undergo uniform acceleration $a$ (recast as $A_I$ and $C_I$ in Rindler region I), while the central node $B$ remains completely stationary. Due to the distinct positions of the accelerated detectors within each bipartite state, the two links experience different relativistic decoherence channels: (i)
for the first link $A$-$B$, detector $A$ undergoes uniform acceleration while detector $B$ remains stationary. Following the derivation in Sec.~II, the final reduced density matrix $\rho_{A_I B}$ in the Bloch-Fano representation takes the same structure as Eq.\eqref{LS5}, where the matrix elements $\alpha_1$, $\beta_1$, and $\gamma_1$ are defined by Eqs.\eqref{l6}-\eqref{lll6}, and the non-zero Bloch correlation coefficients are presented below Eq.\eqref{LS5}; (ii) for the second link $B$-$C$, detector $B$ remains stationary while detector $C$ undergoes uniform acceleration. According to the derivation in Appendix~A, the reduced density matrix $\rho_{B C_I}$ takes the corresponding Bloch-Fano form given in Eq.\eqref{LS6}, with parameters $\alpha_2$, $\beta_2$, and $\gamma_2$ defined by Eqs.\eqref{l6_BC}-\eqref{lll6_BC}, and Bloch coefficients are presented below Eq.\eqref{LS6}.
The nonlocal correlation generated across the network is then quantified by the maximal non-bilocality parameter $\mathcal{B}_{\text{biloc}}^{\max}(\rho_{A_I B} \otimes \rho_{B C_I})$.

\begin{figure}
\begin{minipage}[t]{0.5\linewidth}
\centering
\includegraphics[width=3.0in,height=6.24cm]{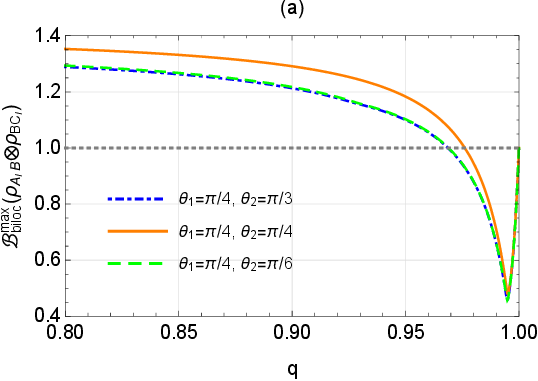}
\label{fig4a}
\end{minipage}%
\begin{minipage}[t]{0.5\linewidth}
\centering
\includegraphics[width=3.0in,height=6.24cm]{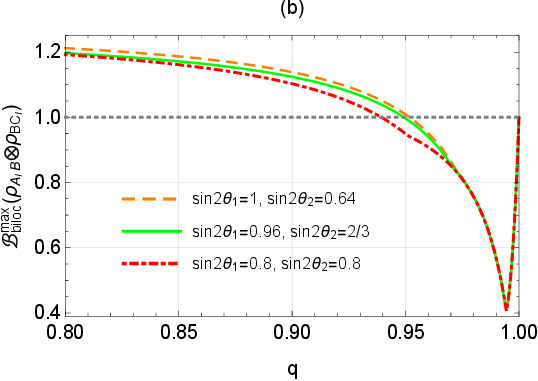}
\label{fig4b}
\end{minipage}%
\caption{The maximum non-bilocality parameter $\mathcal{B}_{\text{biloc}}^{\max}(\rho_{A_I B} \otimes \rho_{B C_I})$ as a function of the acceleration parameter $q$ for different network states. The effective coupling parameter is fixed as $\nu^2=0.01$.}
\label{Fig4}
\end{figure}

In Fig.\ref{Fig4}(a), the maximum non-bilocality parameter $\mathcal{B}_{\text{biloc}}^{\max}(\rho_{A_I B} \otimes \rho_{B C_I})$ is plotted as a function of the acceleration parameter $q$. Here, the initial state $\rho_{A B}$ is fixed as a maximally entangled state ($\theta_1=\frac{\pi}{4}$), while $\rho_{B C}$ is selected from three different states with parameters $\theta_2=\frac{\pi}{3}, \frac{\pi}{4}$, and $\frac{\pi}{6}$. 
As shown in Fig.\ref{Fig4}(a), $\mathcal{B}_{\text{biloc}}^{\max}$ degrades monotonically across all state configurations with increasing acceleration parameter $q$. This trend highlights the destructive nature of the Unruh effect, where the acceleration-induced thermal noise acts as a decoherence channel that disrupts the quantum correlations shared across the network nodes. Remarkably, the curves for $\theta_2 = \frac{\pi}{3}$ and $\theta_2 = \frac{\pi}{6}$ overlap perfectly over the entire range of $q$. This behavior stems from an underlying physical symmetry: the initial entanglement of the state is uniquely determined by the magnitude of $\sin 2\theta_2$. Since $\sin\frac{2\pi}{3} = \sin\frac{\pi}{3}$, both configurations possess identical initial quantum resources. Furthermore, as $q \rightarrow 1$ (the infinite acceleration limit), all curves drop below $1$ ($\mathcal{B}_{\text{biloc}}^{\max} < 1$) due to the severe suppression of quantum coherence, but abruptly recover to exactly $1$ at $q = 1$. At this extreme limit, the infinite-temperature Unruh bath completely eliminates the quantum coherence of the system, transforming the network states into purely classically correlated mixtures.

In Fig.\ref{Fig4}(b), the maximum non-bilocality parameter $\mathcal{B}_{\text{biloc}}^{\max}(\rho_{A_I B} \otimes \rho_{B C_I})$ is plotted as a function of the acceleration parameter $q$ for three different initial network states. To evaluate the optimal resource allocation strategy for sustaining network nonlocality against relativistic degradation, the product of the initial entanglement magnitudes is strictly fixed as $\sin2\theta_1 \cdot \sin2\theta_2 = 0.64$. A clear performance hierarchy emerges among the different distributions of quantum resources: the highly asymmetric configuration, where the first link is maximally entangled ($\sin 2\theta_1 = 1$, $\sin 2\theta_2 = 0.64$), consistently preserves a higher degree of non-bilocality compared to the symmetric distribution ($\sin2\theta_1 = \sin2\theta_2 = 0.8$). This confirms the physical insight that concentrating quantum resources into a single link provides a superior buffer against relativistic degradation compared to distributing them evenly. Specifically, maximizing one link shields its dominant correlation components from premature suppression, thereby maintaining the global network correlation.

\section{Evolution of nonlocality in $n$-locality scenario}
For a general network $\Xi_{m,n}$ consisting of $m$ parties and $n$ sources, which can be regarded as a $k$-independent network, its non-$n$-local correlation can be witnessed by the violation of Ineq.(\ref{eq_B_ineq}). When the number of resources $n$ is fixed, the relativistic degradation of non-$n$-locality under the Unruh-DeWitt detector model depends not only on the acceleration parameter $q$ and coupling strength $\nu^2$, but also on the number of parties $m$ and the independent number $k$.

To evaluate the relativistic evolution of the network under uniform acceleration, we assume that the initial resource states $|\Psi_{A^{i}A^{i+1}}\rangle$ ($i = 1, 2, \dots, n$) distributed on the network are generated from the following family of two-qubit states
\begin{equation}\label{aHq7}
|\Psi_{A^{i}A^{i+1}}\rangle = \sin{\theta_i}|01\rangle + \cos{\theta_i}|10\rangle,
\end{equation}
where $\theta_i \in [0, \frac{\pi}{2}]$.
We consider the chain and star network scenarios illustrated in Fig.\ref{Fig2} and Fig.\ref{Fig3}, respectively. Assuming that all initial resource states are maximally entangled states ($\theta_1=\theta_2=\dots=\theta_i=\dots=\theta_n=\frac{\pi}{4}$), the initial maximum network nonlocality parameters according to Eqs.(\ref{eq6}) and (\ref{eq7}) are unified as $\mathcal{B}_{\text{chain}}^{\max} = \mathcal{B}_{\text{star}}^{\max} = \sqrt{2}$. Obviously, both chain and star networks can generate non-$n$-local correlations initially. In the following, we investigate whether the non-$n$-local correlations of these two networks can persist under relativistic motion.

\subsection{Scenario 1: The chain network under relativistic motion}
First, we consider the chain network scenario with $n$ sources and $n+1$ parties, where the detectors $\mathcal{A}_i$ undergo uniform acceleration $a$ only when $i$ is an odd integer ($i=1,3,5,\dots$), while the detectors at the even nodes remain inertial and stationary. Under this setup, the network state $\rho_{A^{1}A^{2}}\otimes\rho_{A^{2}A^{3}}\otimes\cdots\otimes\rho_{A^{n}A^{n+1}}$ evolves alternately depending on the link index $i$:
\begin{itemize}
\item For any odd-indexed link $i$, the first particle $A^i$ undergoes uniform acceleration while $A^{i+1}$ remains stationary. Therefore, its final density matrix is described by $\rho^{A_I^i A^{i+1}}_t$, where the matrix elements $\alpha_1^{(i)}$, $\beta_1^{(i)}$, and $\gamma_1^{(i)}$ follow the derivation of Sec.~II [Eqs.\eqref{l6}-\eqref{lll6} with $\theta_1 \to \theta_i$].
\item For any even-indexed link $i$, the first particle $A^i$ remains stationary while the second particle $A^{i+1}$ undergoes uniform acceleration. Accordingly, its final density matrix is given by $\rho^{A^i A_I^{i+1}}_t$, with matrix elements $\alpha_2^{(i)}$, $\beta_2^{(i)}$, and $\gamma_2^{(i)}$ following Appendix~A [Eqs.\eqref{l6_BC}-\eqref{lll6_BC} with $\theta_2 \to \theta_i$].
\end{itemize}
When the initial resource states are maximally entangled ($\theta_i = \pi/4$), the symmetry of the state ensures that the correlation matrices remain diagonal. The non-zero Bloch correlation coefficients for both $\rho^{A_I^i A^{i+1}}_t$ and $\rho^{A^i A_I^{i+1}}_t$ are obtained as $t_1^{(i)} = t_2^{(i)} = \frac{2(1-q)}{2(1-q) + \nu^2(1+q)}$ and $t_3^{(i)} = \frac{ \nu^2(1+q)-2(1-q)}{2(1-q) + \nu^2(1+q)}$. Consequently, the positive eigenvalues of the matrix $\sqrt{T^\dagger T}$ required for evaluating $\mathcal{B}_{\text{chain}}^{\rm{max}}$ in Eqs.~\eqref{eq6} and \eqref{eq7} are given by the absolute values of these coefficients, yielding $\delta_j^{(i)} = |t_j^{(i)}|$ ($j=1,2,3$). Due to the complexity of the final expression for $\mathcal{B}_{\text{chain}}^{\rm{max}}$, we refrain from presenting it explicitly here.

\begin{figure}
\centering
\includegraphics[width=3.5in,height=6.5cm]{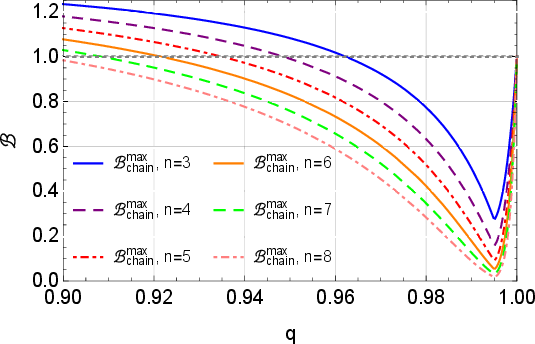}
\caption{$\mathcal{B}_{\text{chain}}^{\max}(\rho_{A_I^1 A^2} \otimes \rho_{A^2 A_I^3} \otimes \cdots \otimes \rho_{A_I^n A^{n+1}})$ as a function of the acceleration parameter $q$ for various network sizes $n$. Initial resource states $\rho_{A^i A^{i+1}}\ (i=1,2,\dots, n)$ are all maximally entangled states. The effective coupling parameter is fixed as $\nu^2=0.01$.}
\label{Fig5}
\end{figure}

In Fig.\ref{Fig5}, we plot the maximum network nonlocality parameter $\mathcal{B}_{\text{chain}}^{\max}$ as a function of the acceleration parameter $q$ for various network sizes $n$.
As shown in Fig.\ref{Fig5}, $\mathcal{B}_{\text{chain}}^{\max}$ monotonically decreases with increasing acceleration parameter $q$, illustrating that the acceleration-induced Unruh thermal noise acts as a decoherence channel that suppresses the global network nonlocality. Crucially, as the number of links $n$ increases, $\mathcal{B}_{\text{chain}}^{\max}$ decays more rapidly, leading to an earlier onset of the ``sudden death'' of non-$n$-locality. This behavior implies that chain networks with a larger number of intermediate nodes exhibit higher sensitivity and lower resilience to relativistic Unruh decoherence.

\subsection{Scenario 2: The star network under relativistic motion}
Next, we turn to the star network scenario comprising a central intermediate node $\mathcal{A}_1$ and $n$ peripheral edge nodes $\mathcal{A}_{i+1}$ ($i=1,2,\dots,n$). We investigate the configuration where the central node $\mathcal{A}_1$ remains completely inertial and stationary, while all the peripheral edge nodes undergo identical uniform acceleration $a$. Under this setup, every single link $\rho_{A^1 A^{i+1}}$ in the star network consistently involves a stationary first qubit and an accelerated second qubit. Therefore, all $n$ links evolve uniformly according to the density matrix $\rho^{A^1 A_I^{i+1}}_t$. 
For initially maximally entangled states, the eigenvalues for each link are again equal to $t_1^{(i)}$, $t_2^{(i)}$ and $t_3^{(i)}$. Due to the complexity of the expression for $\mathcal{B}_{\text{star}}^{\max}$, we refrain from presenting it explicitly here.

\begin{figure}
\centering
\includegraphics[width=3.5in,height=6.5cm]{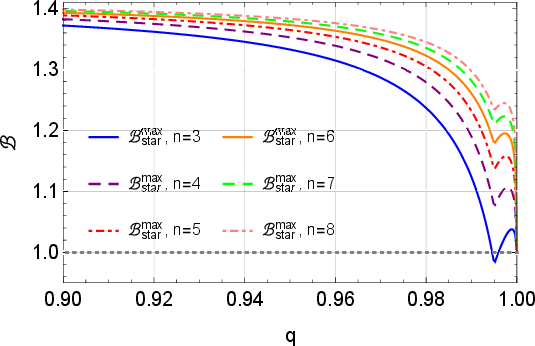}
\caption{$\mathcal{B}_{\text{star}}^{\max}(\rho_{A^1 A_I^2} \otimes \rho_{A^1 A_I^3} \otimes \cdots \otimes \rho_{A^1 A_I^{n+1}})$ as a function of the acceleration parameter $q$ for various network sizes $n$. Initial resource states $\rho_{A^1 A^{i+1}}\ (i=1,2,\dots, n)$ are all maximally entangled states. The effective coupling parameter is fixed as $\nu^2=0.01$.}
\label{Fig6}
\end{figure}

In Fig.\ref{Fig6}, the maximum network nonlocality parameter $\mathcal{B}_{\text{star}}^{\max}$ is plotted as a function of the acceleration parameter $q$ for various network sizes $n$. As anticipated, the acceleration-induced Unruh thermal noise exerts a detrimental effect on the multipartite correlations, driving an initial downward trend across all curves as $q$ increases. However, a striking and anomalous phenomenon emerges at relatively high accelerations. Specifically, for the minimal network size of $n=3$, $\mathcal{B}_{\text{star}}^{\max}$ drops below the classical locality threshold of $1.0$, signifying the ``sudden death'' of the network nonlocality. Remarkably, as $q$ continues to increase, $\mathcal{B}_{\text{star}}^{\max}$ resurges above the classical bound. This non-trivial revival explicitly demonstrates that the vanished network nonlocality can successfully return and undergo a phenomenon of ``rebirth'' under relativistic motion. Moreover, for larger networks ($n > 3$), $\mathcal{B}_{\text{star}}^{\max}$ remains strictly greater than $1.0$ over the entire range of $q$, indicating that star networks with more than three marginal nodes can persistently sustain non-$n$-local correlations regardless of the acceleration magnitude. Furthermore, the maximum nonlocality parameter $\mathcal{B}_{\text{star}}^{\max}$ scales monotonically with the network size $n$. This feature implies that increasing the number of marginal nodes effectively scales up the collective quantum resources shared across the network, thereby significantly bolstering the resilience of the star topology against relativistic Unruh decoherence. In summary, the star network exhibits a significant structural advantage over the chain network in resisting relativistic degradation. While the chain network's nonlocality is highly fragile and scales down rapidly with the addition of intermediate nodes, the star network possesses a unique resilience, capable of driving the revival of vanished network nonlocality under extreme Unruh decoherence.

\section{Conclusions}
In this work, we have investigated the relativistic dynamics of network nonlocality in chain and star quantum networks within the Unruh-DeWitt detector framework. Our results reveal that the response of network nonlocality to relativistic acceleration depends strongly on the underlying network topology.
For the elementary three-node network, we first find that the robustness of nonbilocal correlations can be enhanced by an asymmetric distribution of the initial entanglement resources. When the total amount of initial entanglement is fixed, concentrating a larger fraction of the resource in one link preserves stronger network nonlocality than an approximately symmetric distribution. This result indicates that the allocation of quantum resources among different links provides an additional means of mitigating relativistic degradation.
Notably, for the minimal network with $n=3$, the maximal nonlocality parameter $\mathcal{B}_{\mathrm{star}}^{\max}$ first decreases below the classical threshold and subsequently reenters the non-$n$-local regime at sufficiently large acceleration. This sudden-death--sudden-birth behavior demonstrates that relativistic acceleration does not necessarily lead to a monotonic loss of network nonlocality. Instead, the Unruh effect can suppress network nonlocality over an intermediate acceleration range while allowing its recovery at higher acceleration. This reentrant behavior highlights the important role of network structure in determining the response of collective correlations to relativistic effects.

We further show that chain and star networks exhibit qualitatively different relativistic behaviors. For chain networks, the maximal nonlocality parameter $\mathcal{B}_{\text{chain}}^{\max}$ decreases monotonically with increasing acceleration parameter $q$. Moreover, the degradation becomes increasingly pronounced as the number of links $n$ increases, resulting in an irreversible sudden death of non-$n$-local correlations. Thus, the accumulation of intermediate nodes makes the chain topology progressively more vulnerable to Unruh-induced decoherence. In contrast, the robustness becomes even more pronounced for larger star networks. For $n>3$, $\mathcal{B}_{\text{star}}^{\max}$ remains above the classical threshold throughout the investigated acceleration regime, indicating that non-$n$-local correlations can persist even under strong relativistic acceleration. Moreover, increasing the number of peripheral nodes enhances the maximal network nonlocality and thereby improves the resilience of the star topology against Unruh-induced decoherence. These results establish a clear topology-dependent distinction: increasing the network size tends to amplify the fragility of chain networks, whereas it can enhance the robustness of star networks.

Overall, our results demonstrate that the relativistic degradation of quantum network nonlocality is governed not only by the strength of acceleration and the amount of initial entanglement, but also by the distribution of quantum resources and the underlying network structure. The emergence of reentrant nonlocality in the minimal star network and the persistence of non-$n$-local correlations in larger star networks suggest that network topology can serve as a resource for mitigating relativistic decoherence. These insights provide valuable conceptual guidance for the structural optimization and design of acceleration-resilient architectures for relativistic quantum communication protocols, holding significant potential for long-baseline entanglement-assisted nonlocal phase sensing networks \cite{Stas2026Nature}.

\begin{acknowledgments}
This work is supported by the National Natural Science Foundation of China (Grant No. 12575056) and LiaoNing Revitalization Talents Program (XLYC2503099).	
\end{acknowledgments}


\appendix
\onecolumngrid
\section{Evolution of the system $BC$ under Unruh-DeWitt detector model }
In this appendix, we present the detailed derivation of state evolution for the system $BC$ under the Unruh-DeWitt detector model. We assume that Charlie's detector undergoes uniform acceleration while Bob's detector remains stationary.
Initially, Bob and Charlie share an entangled state in Minkowski spacetime, described by
\begin{equation}\label{AL7}
|\Psi^{BC\phi}_{t_{0}}\rangle=|\Psi_{BC}\rangle\otimes|0_{M}\rangle,
\end{equation}
where $|\Psi_{BC}\rangle$ is defined identically to Eq.(\ref{ql7}).
The total Hamiltonian of the system $BC\phi$ is given by
\begin{equation}\label{AL9}
H_{BC\phi}=H_{B}+H_{C}+H_{KG}+H_{\rm{int}}^{C\phi},
\end{equation}
where $H_{T} = \Omega T^{\dagger}T$ ($T=B,C$) denotes the free Hamiltonian of the detectors with energy gap $\Omega$, $H_{KG}$ is the Hamiltonian of the massless scalar field, and  $H_{\rm{int}}^{C\phi}$ describes the coupling between Charlie's detector and the field. This localized interaction is modeled via the Unruh-DeWitt coupling around the detector $C$ as
\begin{equation}\label{AL8}
H_{\rm{int}}^{C\phi}(t)=\epsilon(t)\int_{\Sigma_{t}}d^{3} \mathbf{x}\sqrt{-g}\phi(x)[\chi(\mathbf{x})C+\bar{\chi}(\mathbf{x})C^{\dagger}].
\end{equation}

In the weak-coupling regime, the final state $|\Psi^{BC\phi}_{t}\rangle$ of the atom-field system at time $t=t_{0}+\Delta$ can be calculated in first-order perturbation theory over the coupling constant $\epsilon$.  Under the evolution generated by the Hamiltonian $H_{BC\phi}$ in Eq.(\ref{AL9}), the final state $|\Psi^{BC\phi}_{t}\rangle$ at time $t$ takes the form
\begin{equation}\label{qqL1}
|\Psi^{BC\phi}_{t}\rangle=\{I-i[\phi(f)C+\phi(f)^{\dagger}C^{\dagger}]\}|\Psi^{BC\phi}_{t_{0}}\rangle,
\end{equation}
where the smeared field operator $\phi(f)$ retains the same definition as in Eq.(\ref{qq2}).

Substituting the initial state of system $BC\phi$ in Eq.(\ref{AL7}) into Eq.(\ref{qqL1}), we obtain the evolved state for the system $BC$ in terms of the Rindler operators as
\begin{equation}\label{qqqL2}
|\Psi^{BC\phi}_{t}\rangle=|\Psi^{BC\phi}_{t_{0}}\rangle+\sin{\theta_2}|00\rangle\otimes \big[a^{\rm \dagger}_{RI}(\lambda)|0_{M}\rangle \big]+\cos{\theta_2}|11\rangle\otimes \big[a_{RI}(\bar{\lambda})|0_{M}\rangle \big].
\end{equation}
By applying the Bogoliubov transformations outlined in Eqs.(\ref{qq3}) and (\ref{qq5}), the evolved state in Eq.(\ref{qqqL2}) can be recast as
\begin{equation}\label{qqqL5}
|\Psi^{BC\phi}_{t}\rangle=|\Psi^{BC\phi}_{t_{0}}\rangle+\nu \left [ \sin{\theta_2}\frac{|00\rangle
\otimes|1_{\tilde{F}_{1\Omega}}\rangle}{(1-e^{-2\pi\Omega/a})^{1/2}} + e^{-\pi\Omega/a} \cos{\theta_2} \frac{|11\rangle\otimes|1_{\tilde{F}_{2\Omega}}\rangle}{(1-e^{-2\pi\Omega/a})^{1/2}} \right],
\end{equation}
where $\tilde{F}_{i\Omega}=F_{i\Omega}/\nu$ ($i=1,2$).  

To determine the state of the detectors after their interaction with the field, we trace out the external field degrees of freedom. This yields the reduced density matrix of the system $BC$:
\begin{equation}
\rho^{{BC}}_{t}=\|\Psi^{BC\phi}_{t}\|^{-2}{\mathrm{Tr}}_{\phi}|\Psi^{BC\phi}_{t}\rangle\langle \Psi^{BC\phi}_{t}|,
\end{equation}
where the normalization factor $\|\Psi^{BC\phi}_{t}\|^{2}$ is calculated as 
\[
\|\Psi^{BC\phi}_{t}\|^{2}=1+\frac{\nu^{2}(\sin^2\theta_2+e^{-2\pi\Omega/a} \cos^2\theta_2)}{1-e^{-2\pi\Omega/a}}.
\]
Accordingly, the reduced density matrix of the Bob-Charlie detector subsystem takes the following form:
\begin{equation}\label{ppL6}
\rho^{BC}_{t}=
 \left(\!\!\begin{array}{cccc}
{\hskip 6pt}\gamma_2&{\hskip 6pt} 0& 0&{\hskip 6pt} 0{\hskip 6pt}\\
{\hskip 6pt}0&{\hskip 6pt} 2 \alpha_2 \sin^2\theta_2& \alpha_2 \sin2\theta_2&{\hskip 6pt} 0{\hskip 6pt}\\
{\hskip 6pt}0&{\hskip 6pt} \alpha_2 \sin2\theta_2& 2 \alpha_2 \cos^2\theta_2&{\hskip 6pt} 0{\hskip 6pt}\\
{\hskip 6pt}0&{\hskip 6pt} 0& 0&{\hskip 6pt} \beta_2{\hskip 6pt}
\end{array}\!\!\right) ,
\end{equation}
where the matrix elements $\alpha_2$, $\beta_2$, $\gamma_2$ are given by 
\begin{equation}\label{l6_BC}
\alpha_2=\frac{1-q}{2(1-q)+2\nu^{2}(\sin^2\theta_2+q \cos^2\theta_2)},
\end{equation}
\begin{equation}\label{ll6_BC}
\beta_2=\frac{\nu^{2} q \cos^2\theta_2}{(1-q)+\nu^{2}(\sin^2\theta_2+q \cos^2\theta_2)},
\end{equation}
\begin{equation}\label{lll6_BC}
\gamma_2=\frac{\nu^{2} \sin^2\theta_2}{(1-q)+\nu^{2}(\sin^2\theta_2+q \cos^2\theta_2)}.
\end{equation}
Here, the acceleration parameter $q \equiv e^{-2\pi\Omega/a}$ , the effective coupling strength $\nu$, and the weak-coupling condition $\nu^2 \ll 1$ retain the identical physical meanings and validity parameters established in the main text.

From Eq.(\ref{ppL6}), it is obvious that the reduced density matrix of the detectors $\rho^{{BC}}_{t}$ satisfies the form of the bipartite $X$-state. Rewriting $\rho^{{BC}}_{t}$ in the Bloch-Fano decomposition according to Eq.(\ref{LS1}) yields
\begin{equation}\label{LS6}
\begin{split}
\rho^{BC}_{t} =& \frac{1}{4} \Big[ I_2 \otimes I_2 + (\gamma_2 - 2 \alpha_2 \cos2\theta_2 - \beta_2) \sigma_3 \otimes I_2 + (\gamma_2 + 2 \alpha_2 \cos2\theta_2 - \beta_2) I_2 \otimes \sigma_3 \\
& + 2 \alpha_2 \sin2\theta_2 \sigma_1 \otimes \sigma_1 + 2 \alpha_2 \sin2\theta_2 \sigma_2 \otimes \sigma_2 + (\gamma_2 - 2 \alpha_2 + \beta_2) \sigma_3 \otimes \sigma_3 \Big].
\end{split}
\end{equation}
Therefore, the corresponding components can be defined  as $t_1^{(2)}=t_2^{(2)}=2 \alpha_2 \sin2\theta_2$, $t_3^{(2)}=\gamma_2 - 2 \alpha_2 + \beta_2$, $a_3^{(2)}=\gamma_2 - 2 \alpha_2 \cos2\theta_2 - \beta_2$ and $b_3^{(2)}=\gamma_2 + 2 \alpha_2 \cos2\theta_2 - \beta_2$.

\end{document}